\documentclass[twocolumn,showpacs,amsmath,amssymb,aps,prb,superscriptaddress]{revtex4-1}

\usepackage[]{cleveref}
\usepackage[]{verbatim}
\usepackage[]{color}
\usepackage[abs]{overpic}
\usepackage{rotating}
\usepackage{tabularx}

\usepackage{amsfonts}
\usepackage{amsmath}
\usepackage{amssymb}
\usepackage{bm}
\usepackage{graphicx}
\usepackage[breaklinks,colorlinks=true,citecolor=blue]{hyperref}
\usepackage{mathrsfs}
\usepackage[lofdepth,lotdepth,caption=false]{subfig}
\usepackage{varwidth}
\usepackage{wrapfig}
\usepackage{times}
\usepackage{longtable}
\usepackage{multirow}
\usepackage{tikz}
\usepackage{colortbl}
\definecolor{darkblue}{RGB}{0 60 120}
\definecolor{eggplant}{RGB}{190 10 150}
\definecolor{darkgray}{RGB}{70 70 70}
\definecolor{lightgray}{RGB}{80 80 80}
\definecolor{lightgray2}{RGB}{245 215 110}
\definecolor{lightgray3}{RGB}{255 0 0}

\newcommand{\rucl}{$\alpha$-RuCl$_3$}

\graphicspath{{./}}

\begin{document}

\title{Crystal structure and magnetism in ${\alpha}$-RuCl$_3$: an {\it \textbf{ab-initio}} study}

\author{Heung-Sik Kim}
\affiliation{Department of Physics and Center for Quantum Materials, 
University of Toronto, 60 St. George St., Toronto, Ontario, M5S 1A7, Canada}

\author{Hae-Young Kee}
\email{hykee@physics.utoronto.ca}
\affiliation{Department of Physics and Center for Quantum Materials, 
University of Toronto, 60 St. George St., Toronto, Ontario, M5S 1A7, Canada}
\affiliation{Canadian Institute for Advanced Research / Quantum Materials Program,
Toronto, Ontario MSG 1Z8, Canada}

\begin{abstract}
\rucl~has been proposed recently as an excellent playground for exploring Kitaev physics on a two-dimensional (2D) honeycomb lattice. However, structural clarification of the compound has not been completed, which is crucial in understanding the physics of this system. Here, using {\it ab-initio} electronic structure calculations, we study a full three dimensional (3D) structure of \rucl~including the effects of spin-orbit coupling (SOC) and electronic correlations. Three major results are as follows; i) SOC suppresses dimerization of Ru atoms, which exists in other Ru compounds such as isostructural Li$_2$RuO$_3$, and making the honeycomb closer to an ideal one. ii) The nearest-neighbor Kitaev exchange interaction between the $j_{\rm eff}$=1/2 pseudospin depends strongly on the Ru-Ru distance and the Cl position, originating from the nature of the edge-sharing geometry. iii) The optimized 3D structure without electronic correlations has $P{\bar 3}1m$ space group symmetry independent of SOC, but including electronic correlation changes the optimized 3D structure to either $C2/m$ or $Cmc2_1$ within 0.1 meV per formula unit (f.u.) energy difference. The reported $P3_112$ structure is also close in energy. The interlayer spin exchange coupling is a few percent of in-plane spin exchange terms, confirming \rucl~is close to a 2D system. We further suggest how to increase the Kitaev term via tensile strain, which sheds new light in realizing Kitaev spin liquid phase in this system.
\end{abstract}
\maketitle


\section{Introduction}
There have been a number of studies on quasi-two-dimensional 
systems having both spin-orbit coupling (SOC) and on-site Coulomb
interactions, which are believed to host unconventional magnetic orders and spin liquid phases
\cite{witczak2014correlated,rau_review}.
One promising candidate is \rucl, where edge-sharing RuCl$_6$ octahedra form two-dimensional
RuCl$_3$ layers in which Ru honeycomb layers reside\cite{Stroganov1957,plumb2014alpha,jasears,Kubota_C2m_XY,Majumder,hskim_RuCl3,luke2,luke,Banerjee}. 
Compared to its
5$d$ transition metal oxide counterparts $\alpha$-$A_2$IrO$_3$ 
($A$=Li,Na)\cite{FYe_NIO,Gretarsson,LIO,singh_Li2IrO3,rau2014generic}, 
\rucl~has closer-to-ideal RuCl$_6$ octahedra\cite{Stroganov1957}, so it was 
proposed as an excellent platform to explore the Kitaev physics and 
related magnetism despite weaker 
SOC\cite{kitaev2006anyons,jackeli2009mott,plumb2014alpha,luke2,Banerjee}. 
A few recent reports suggest the presence of strong Kitaev-type bond-dependent
exchange interactions in \rucl\cite{jasears}, which originate
from the cooperation between the intermediate SOC in Ru atom and the Coulomb 
interaction\cite{hskim_RuCl3}. A zigzag-type magnetic order within the RuCl$_3$
layer is also predicted and observed, which is proximate to the Kitaev
spin-liquid phase\cite{hskim_RuCl3,jasears}. 

\begin{figure}
  \centering
  \includegraphics[width=0.40 \textwidth]{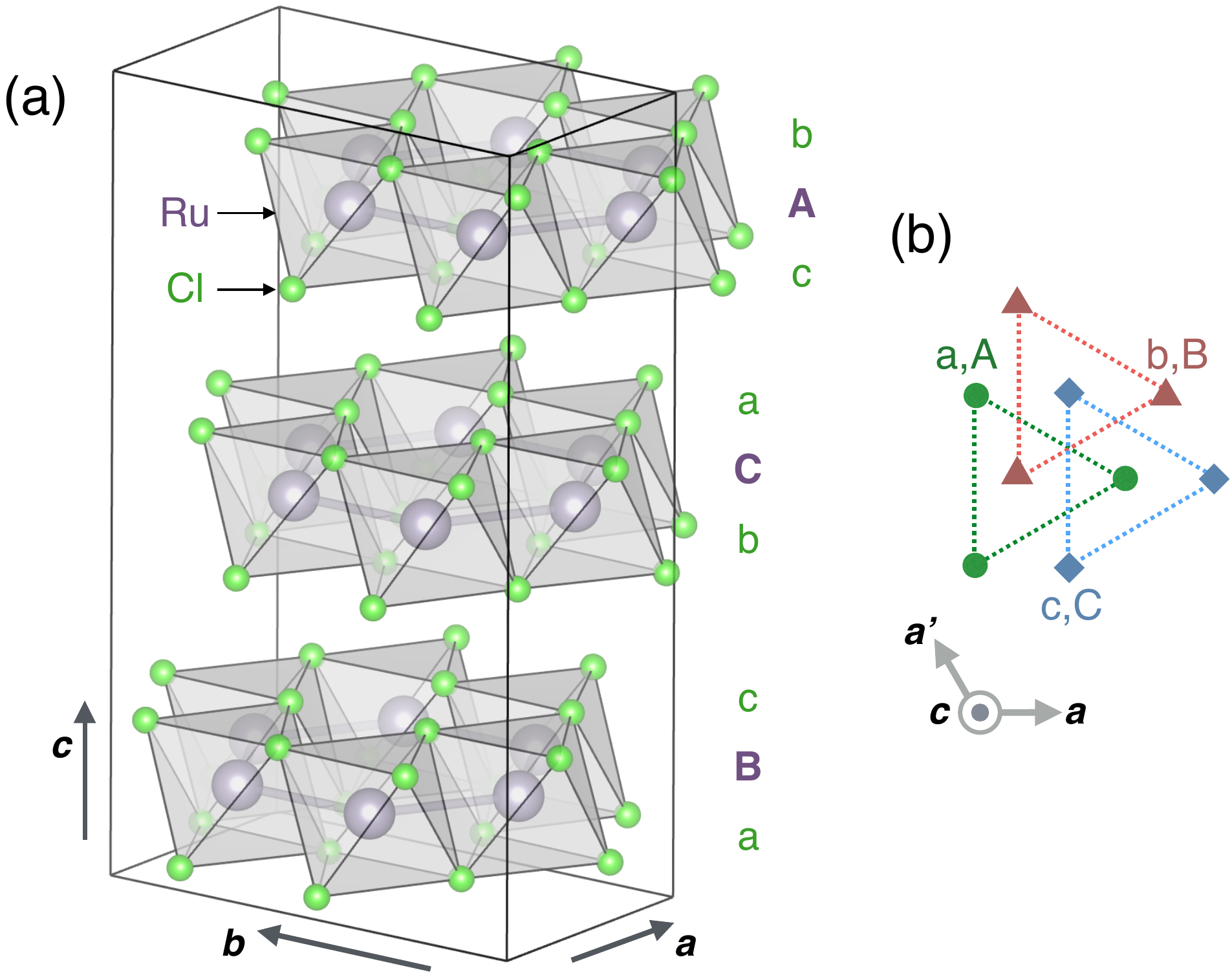}
  \caption{(Color online)
  (a) Crystal structure of \rucl with $C2/m$ space group. 
  Solid lines depict a monoclinic unit cell. 
  (b) Schematic view of three triangular sublattices on which
  Ru and Cl layers are located. Stacking indices for
  Ru honeycomb and Cl triangular layers are shown on the right 
  side of (a), where indices for Ru and Cl layers are expressed as
  capital and lowercase letters respectively.
  }
  \label{fig:struct}
\end{figure}

In previous studies \rucl~was considered as a two-dimensional system
with an ideal Ru honeycomb lattice, but such assumption needs further clarification. 
A potential Ru layer distortion, which is observed in an isostructural compound 
Li$_2$RuO$_3$\cite{LRO_1,LRO_2}, might happen in this compound. Furthermore, 
\rucl~has a three-dimensional crystal structure consisting of RuCl$_3$ layer 
stacking, and interlayer coupling and interaction terms can introduce another 
complication. Experimentally, both
$P3_112$ and $C2/m$ space groups have been suggested as the crystalline symmetry 
in this compound\cite{Stroganov1957,fletcher1967x,RuCl3_C2m,Kubota_C2m_XY,Banerjee}. 
As an illustrative
example, Fig. \ref{fig:struct}(a) shows the crystal structure of \rucl~with a
$C2/m$ space group symmetry, where adjacent RuCl$_3$ layers within the unit 
cell is related to each other by a translation along the $a$-axis in the figure.
Stacking faults can easily be introduced in this layered structure as in the 
case of $\alpha$-$A_2$IrO$_3$\cite{choi2012spin}, which obscures
further clarification of the crystal structure. 
Effect of interlayer exchange interactions from the layer stacking 
on the ground state magnetic properties of this system is 
not well understood either. More interestingly, a sample-dependent two-transition 
behavior is reported, where two different magnetic order peaks at $T_{N1} \simeq 14$ K 
and $T_{N2} \simeq 8$ K with two- and three-layer 
$c$-axis periodicity, respectively, are observed in neutron diffraction measurement
\cite{Banerjee}. These issues pose a question on the relation between crystal structure 
and magnetism in this system. 

In pursuit of such motivations, in this work we perform {\it ab-initio}
calculations for the structural properties of \rucl~ and their impact on magnetism. 
We present three main results; i) Role of SOC and zigzag magnetic order on 
the single-layer RuCl$_3$ structure is discussed. We found that SOC prefers ideal
honeycomb lattice by preventing Ru-Ru dimer formations, and the presence of
in-plane zigzag order tends to give small monoclinic distortion commensurate
to the magnetic order. ii) Effect of Ru-Cl and Ru-Ru distance to 
the exchange interactions and magnetism is discussed, where the hopping channels
within the nearest-neighbor(NN) Ru $t_{\rm 2g}$ orbitals and the resulting exchange 
interactions between the SOC-induced $j_{\rm eff}$=1/2 pseudospins strongly depend 
on the Ru-Cl and Ru-Ru distance. Such behavior originates from the 
existence of multiple hopping channels in the $t_{\rm 2g}$ orbitals, 
which enables {\it 'leveraging'} magnetism with rather small amount of 
structural changes. iii) Stability of crystal structures with 
different stacking orders is discussed by comparing relative total energies. 
We have found that, structures with $C2/m$\cite{RuCl3_C2m} and $Cmc2_1$ space 
group symmetries are most favorable with almost degenerate energies. Previously 
suggested $P3_112$ structure\cite{Stroganov1957,fletcher1967x}
yields total energy comparable to those of $C2/m$ and $Cmc2_1$ structures
with the energy difference smaller than 0.4 meV per formula unit (f.u.). 
Energy differences between different interlayer magnetic orders are smaller 
than 0.1 meV / f.u., and magnitude of interlayer exchange interactions
estimated from interlayer hopping integrals are smaller than 0.05 meV.
These observations justify the 
employment of two-dimensional spin models in exploring magnetism in \rucl.
We further propose how to increase the Kitaev term using tensile strain
or uniaxial pressure to realize the Kitaev spin liquid phase. 


This manuscript is organized as follows. After showing computational 
details in Sec. \ref{sec:detail}, structural properties of single-layer RuCl$_3$ 
and its relation to magnetism is presented in Sec. \ref{sec:1L}. 
The effect of SOC and zigzag magnetic order to the single-layer RuCl$_3$ structure, and
the relation between the structure and magnetism are discussed in 
Sec. \ref{subsec:Ru} and Sec. \ref{subsec:Cl}, respectively. 
In Sec. \ref{sec:23LwoSOC} and \ref{sec:23LwSOCU}, results on the
stacking without and with the Coulomb interaction and magnetism
are shown, respectively. Summary and conclusion follow in Sec. \ref{sec:sum}. 

\section{Computational details}
\label{sec:detail}
For the electronic structure calculations, we employed 
the Vienna {\it ab-initio} Simulation Package (VASP), which uses the 
projector-augmented wave (PAW) basis set\cite{VASP1,VASP2}. 
370 eV of plane wave energy cutoff was used, and for
$k$-point sampling 15$\times$15 and 8$\times$6$\times$4(6) Monkhorst-Pack grid 
were adopted for single-layer primitive cell and monoclinic cells with three
(two) layer $c$-axis peroidicity. Tetrahedron method with Bl\"{o}chl correction
was used for the calculation of density of states\cite{Tetra}. 
On-site Coulomb interactions are 
incorporated using the Dudarev's rotationally invariant DFT+$U$ formalism\cite{Dudarev}
with effective $U_{\rm eff} \equiv U-J = 2$ eV. For each configuration
with different unit cell, $U_{\rm eff}$ value, and magnetic order, structural optimization is
performed with a force criterion of 1 meV / \AA. 
Unless specified, a revised Perdew-Burke-Ernzerhof generalized gradient 
approximation (PBEsol)\cite{PBEsol} was used for structural optimization and total energy 
calculations. Note that, PBEsol functional yielded reasonable results for the stacking 
order of bilayer transition metal dichalcogenides in comparison to the van der Waals 
functionals\cite{tmd_vdw}. Results with employing vdW functionals are shown in Appendix. 
Interlayer hopping integrals were obtained by employing maximally-localized Wannier orbital 
(MLWF) formalism\cite{MLWF1,MLWF2} implemented in Wannier90 package\cite{Wannier90}. 
Also, for comparison of the magnetism in the single-layer structures in Sec. \ref{sec:1L},
a linear-combinaion-of-pseudo-atomic-orbital basis code OPENMX\cite{openmx,han2006n}
was used, where double zeta plus polarization (DZP) bases, 500 Ry of energy cutoff for real
space integrations, and the Perdew-Zunger parameterization for the local density approximation 
were employed\cite{CA,PZ}.

\section{Relation between structure and magnetism in RuCl$_3$ single layer}
\label{sec:1L}

In this section, structural changes due to the lattice optimization and 
their effect to the magnetism is discussed in the RuCl$_3$ single layer. 
The initial trial structure we chose is the one reported in Ref. \onlinecite{RuCl3_P3112},
which was used in the Ref. \onlinecite{hskim_RuCl3}. The lattice optimization
gives rise to in-plane structural changes, and here we present the optimized 
structures focusing on the difference from the old one. 
Since we found that such behavior and the resulting changes in
magnetism also occur in the full 3D structures, which are presented in 
Sec. \ref{sec:23LwoSOC} and \ref{sec:23LwSOCU}, below we first discuss the single layer results. 


\subsection{Effect of SOC on in-plane Ru dimerization}
\label{subsec:Ru}

\begin{figure}
  \centering
  \includegraphics[width=0.45 \textwidth]{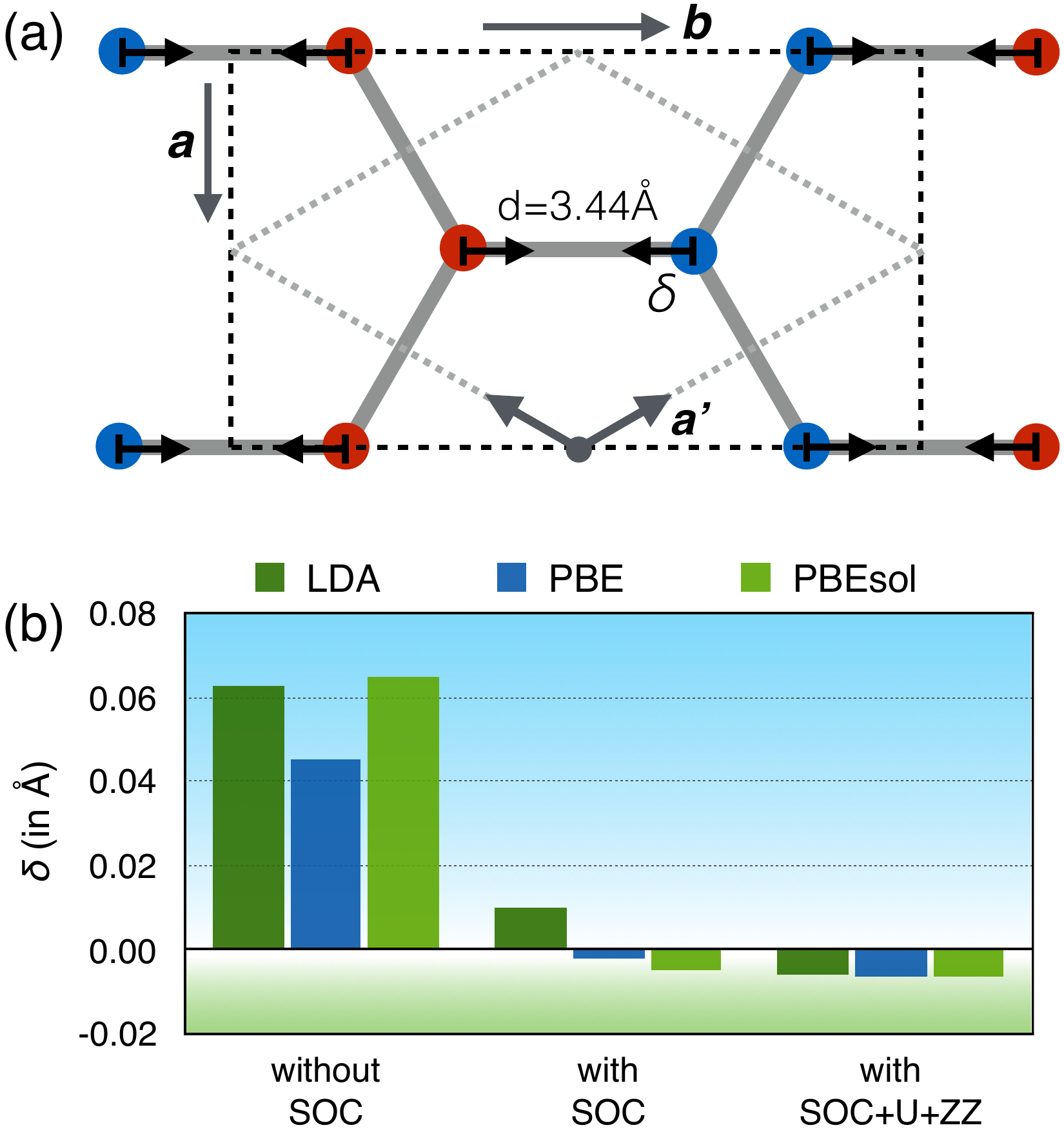}
  \caption{(Color online)
  (a) Schematic figure of Ru honeycomb with colored circles
  depicting Ru sites. Grey dotted and black dashed squares 
  represent the primitive and monoclinic unit cells respectively, 
  where colors on Ru sites show the zigzag magnetic order in a 
  monoclinic unit cell.
  (b) Size of Ru distortion $\delta$ under different exchange-correlations 
  functionals and with/without presence of SOC, $U_{\rm eff}$, and
  in-plane zigzag magnetic order. Note that, positive and negative
  $\delta$ correspond to Ru dimer and zigzag chain formations, respectively. 
  }
  \label{fig:1layer}
\end{figure}

First, the effect of SOC and magnetism with $U_{\rm eff}$ on a Ru honeycomb 
lattice is discussed in this subsection. Fig. \ref{fig:1layer} summarizes the
results, where the sizes of Ru displacements $\delta$ from the ideal 
honeycomb lattice after structural optimizations under different conditions
are shown. Positive and negative values of $\delta$ in Fig. \ref{fig:1layer}(b) 
correspond to Ru dimerization and Ru zigzag chain formation,
respectively, as shown in Fig. \ref{fig:1layer}(a). 
Note that, the lattice constants are fixed to the experimentally observed 
 $a$ = $a_0$ = 5.96\AA~ and $b$=$\sqrt{3}a_0$. 
Without including SOC and Coulomb interactions, 
the two Ru atoms in the unit cell tend to dimerize to lower the energy 
as shown in Fig. \ref{fig:1layer}(a). The presence of dimer formation
is robust against different choice of exchange-correlation functionals 
--- Perdew-Zunger parametrization of local density approximation (LDA) 
\cite{perdew1981self}, PBE\cite{PBE}, and PBEsol --- with slightly different 
size of $\delta$ as shown in Fig. \ref{fig:1layer}(b). Similar dimer formation 
was reported in other layered honeycomb compound Li$_2$RuO$_3$, 
of which origin is suggested to be the $\sigma$-like direct bonding
between the neighboring Ru $t_{\rm 2g}$ orbitals\cite{LRO_1,LRO_2}. 


Since the dimer formation breaks the Ru $t_{\rm 2g}$ degeneracy,
we expect that SOC would not favor the dimer formation. The spin-orbit
entangled $j_{\rm eff}$ orbitals, which emerges under the
presence of cubic crystal fields and SOC\cite{kim2008novel,kim2009phase}, 
does not favor orbital polarization between the $t_{\rm 2g}$ 
--- $d_{xy}$, $d_{xz}$, and $d_{yz}$ --- orbitals. Indeed, structural
optimizations including SOC yield significant reduction of 
dimerization as shown in the middle of Fig. \ref{fig:1layer}(b). 
Although there are small differences between LDA, PBE, and PBEsol 
results, the role of SOC in preventing the dimerization is evident. 
Additionally, inclusion of the on-site Coulomb interaction without
the presence of magnetism is expected to enhance the idealness of 
the Ru honeycomb lattice, since it was shown previously that the on-site 
Coulomb interaction effectively enhances the size of 
SOC\cite{Liu_SRO,hskim_RuCl3}.

Next we show the effect of in-plane zigzag magnetic order, which is
predicted to occur when SOC and the Coulomb interaction are incorporated 
in {\it ab-initio} calculations\cite{hskim_RuCl3} and observed in 
experiments\cite{jasears,Banerjee}. 
Right columns of Fig. \ref{fig:1layer}(b) show the results from 
calculations including SOC, $U_{\rm eff}=2$eV, and the zigzag order. 
The enlarged monoclinic unit cell and the magnetic configuration are
shown in Fig. \ref{fig:1layer}(a), where the red and blue colored 
circles represent Ru sites with antiparallel moments to each other. Regardless 
the choice of functional, $\delta$ shows negative values with almost 
same magnitude. The resulting structure is commensurate to the zigzag 
magnetic order as shown in Fig. \ref{fig:1layer}(a), suggesting a finite
magneto-elastic coupling in this compound. 

\subsection{Effects of Cl displacement and lattice constant change 
to the exchange interactions between the $j_{\rm eff}$=1/2 pseudospins}
\label{subsec:Cl}

\begin{figure}
  \centering
  \includegraphics[width=0.4 \textwidth]{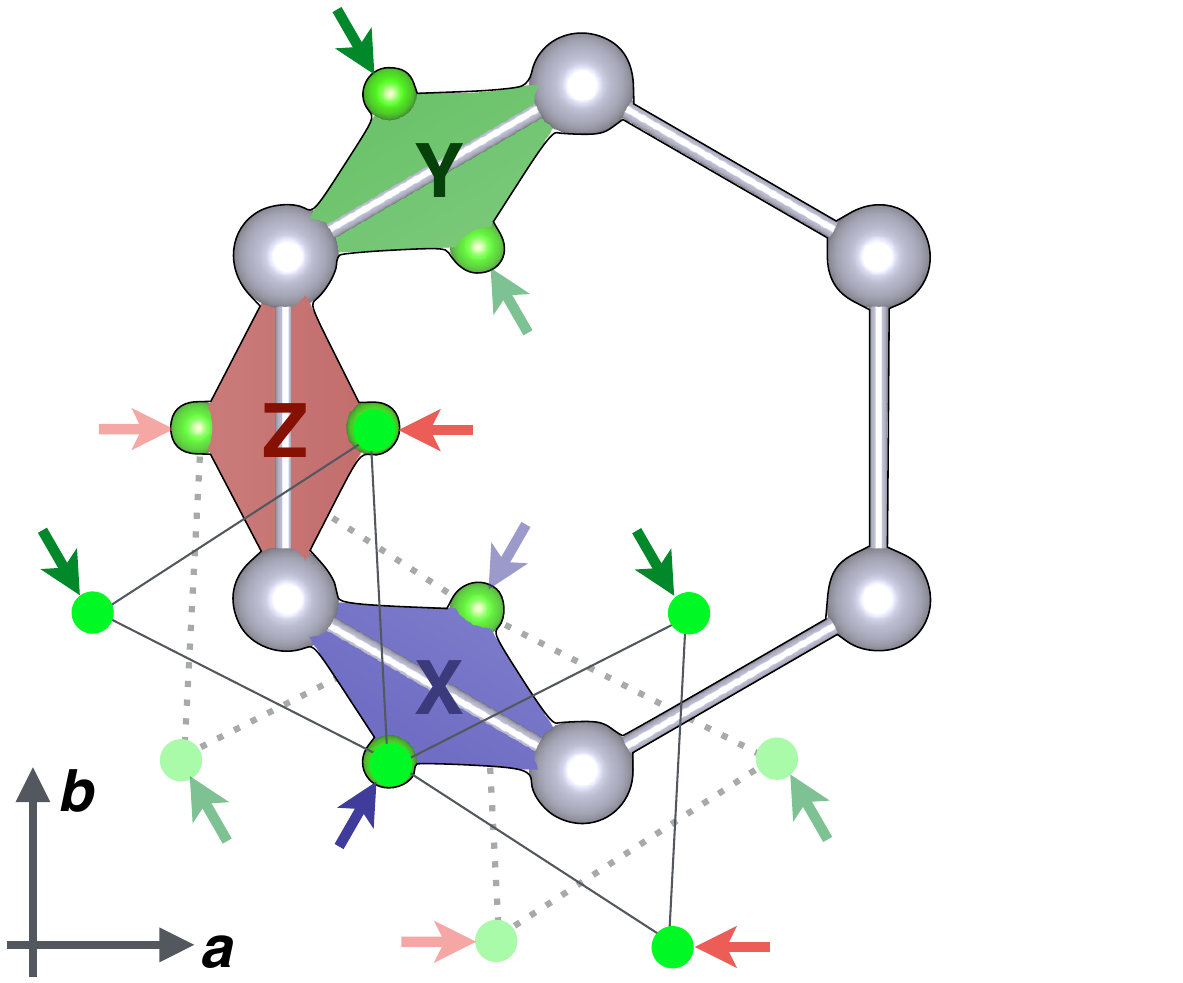}
  \caption{(Color online)
  Schematic figure showing the direction of Cl displacement from the ideal 
  position after the structural optimization. Three inequivalent NN bonds ---
  Z-, X-, and Y- bonds ---  and the displacements of participating Cl atoms 
  therein are depicted by red, blue, and green planes and arrows, respectively. 
  Cl triangles located above and below the Ru plane are represented as solid and 
  dotted triangles, respectively. 
  }
  \label{fig:bond}
\end{figure}

Here we discuss the Cl displacement after the optimization and its impact to the
exchange interactions between the neighboring Ru $j_{\rm eff}$=1/2 pseudospins. 
Fig. \ref{fig:bond} shows the displacement of Cl atoms after structural optimization, where
the two Cl atoms participating in each NN Ru bond move toward the bond 
center. When the in-plane lattice constants are fixed to be $a$ = $a_0$ and
$b$=$\sqrt{3}a_0$, structural optimization with SOC only 
(no $U_{\rm eff}$ and magnetism) yields reduced Cl height of 1.43\AA~ to 1.34\AA~
with respect to the Ru plane, and the Cl triangles above and below Ru plane rotates 
by 2.7$^\circ$ in opposite direction as shown in the figure. The Ru-Cl-Ru NN bond angle 
increases from 89.1$^\circ$ to 93.8$^\circ$. After allowing the lattice constants to relax, 
the lattice constants reduce to $a$ = 0.981$a_0$ and $b$ = 0.986$b_0$ when 
SOC was employed with the monoclinic distortion allowed. With 
$U_{\rm eff}$ = 2 eV and the zigzag magnetic order, they are increased to $a$ = $1.011a_0$
and $b$ = $1.006b_0$. The averaged Ru-Cl distance changes from 2.34\AA~ to 2.36\AA~
in the nonmagnetic calculation with $U_{\rm eff}$ = 0 eV to the magnetic results with
$U_{\rm eff}$ = 2 eV, but both of them are shorter than the distance of 2.45\AA~
in the initial trial structure. Note that, when the monoclinic distortion is allowed, 
the NN Z-bond in Fig. \ref{fig:bond} becomes inequivalent to the X and Y bonds, where
the X and Y bonds form the zigzag chain in Fig. \ref{fig:1layer}(a). Also,
no Ru-Cl bond length disproportionation is observed in all of our results, implying no
Jahn-Teller distortion in this system. 

\begin{figure}
  \centering
  \includegraphics[width=0.45 \textwidth]{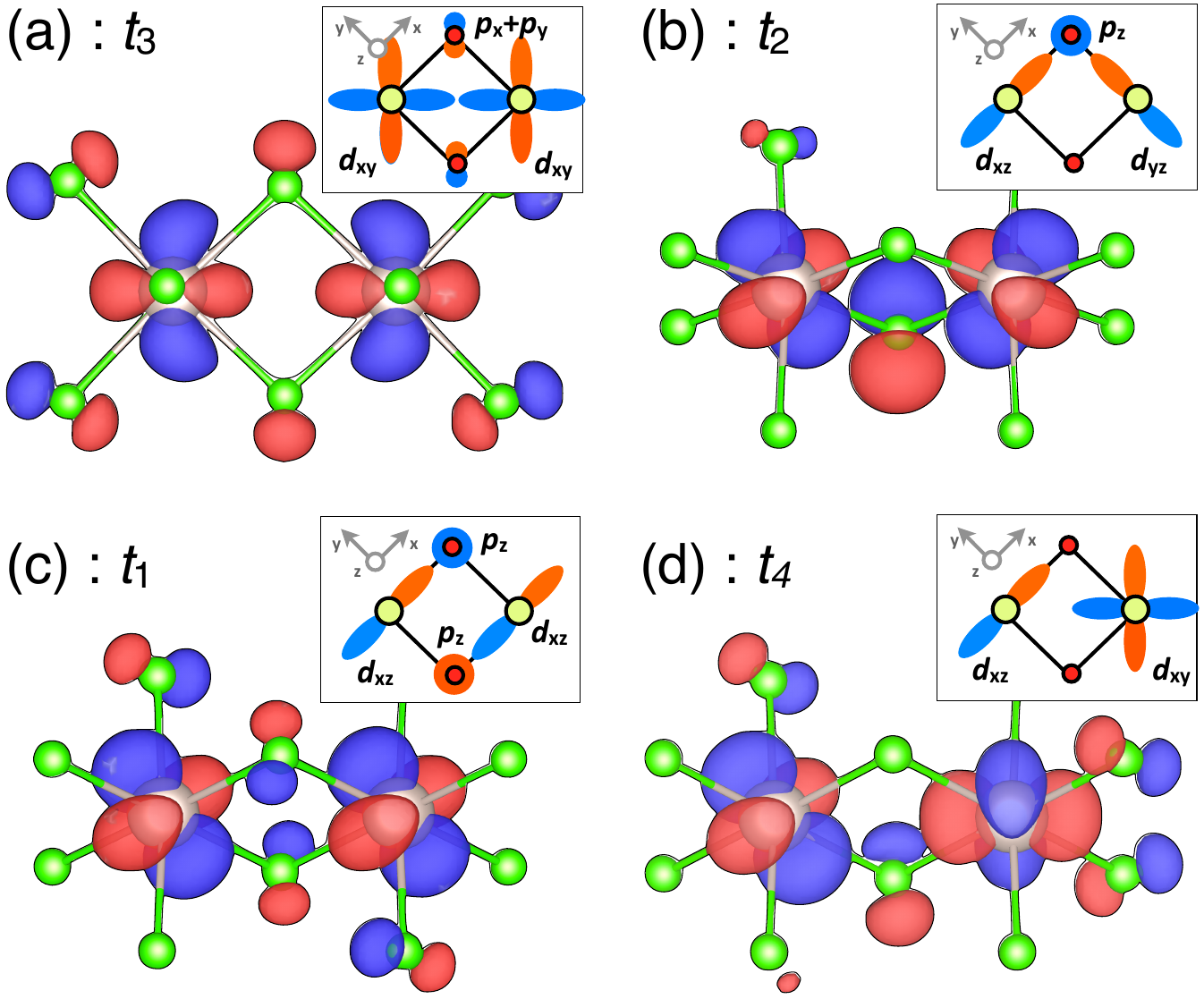}
  \caption{(Color online)
  Four major NN hopping channels --- (a) $t_3$, (b) $t_2$, (c) $t_1$, and (d) $t_4$ ---
  within the $t_{\rm 2g}$ subspace. For each hopping channel, the participating 
  $t_{\rm 2g}$ Wannier orbitals are plotted, where the schematics for each channel
  is represented in the inset. Note that, $t_3$ and $t_2$ terms depends more 
  sensitively to the structural change than $t_1$ and $t_2$. 
  }
  \label{fig:wan}
\end{figure}

Due to the presence of inversion symmetry at the bond center and 
additional trigonal distortion in RuCl$_6$ octahedra, the hopping integrals between the 
NN Ru $t_{2g}$ Wannier orbitals have the following form\cite{rau2014generic,rau_trigonal},
\begin{equation}
\hat{T} = \left(
\begin{array}{ccc}
t_1 & t_2 & t_4 \\
t_2 & t_1 & t'_4 \\
t_4 & t'_4 & t_3
\end{array} \right), \nonumber
\end{equation}
where each hopping channel is displayed in Fig. \ref{fig:wan} with the participating 
Ru $t_{\rm 2g}$ Wannier orbitals therein. As shown in the figure, while $t_1$ originates mainly 
from the $\delta$- and $\sigma$-like $d$-$d$ direct overlap integrals, $t_2$ is mostly 
from the $\pi$-type indirect overlap dominated by $d$-$p$-$d$ hopping between the 
Ru and intervening Cl $p$ orbitals. Note that, $t_3$ channel has both the $d$-$d$ direct 
overlap and $d$-$p$-$d$ indirect overlap which has opposite signs to each other. 
Also, due to the small trigonal distortion, the small $t_4$ and $t'_4$ terms are introduced, 
where the difference between them introduced by the monoclinic distortion is negligibly 
small. 

\begin{table*}
  \centering
  \setlength\extrarowheight{2pt}
  \begin{tabular}{@{\extracolsep{\fill}}lrrrrrrrrrr} 
   \hline\hline   & $d^{\rm avg}_{\rm Ru-Cl}$ & $d_{\rm Ru-Ru}$ &~~~~~~~~~~~~~~~~ $t_1$ & $t_2$
                  & $t_3$ & $t_4$ & ~~~~~~~~~~~~~~~~$J$ & $K$ & $\Gamma$ & $\Gamma'$ \\ 
                  & \multicolumn{2}{r}{(in \AA)}  & \multicolumn{4}{r}{(in eV)} & \multicolumn{4}{r}{(in meV)} \\[4pt] \hline
 {\it ~~Case 0 structure: old $P3_112$ structure}&&&&&&&&&& \\[-2pt]
	\multicolumn{11}{l}
		{\footnotesize\it 
		(from Ref. \onlinecite{Stroganov1957}, $a$=$a_0$, $b$=$b_0$)} \\[3pt]
    NN            & 2.45 & 3.44 & +0.066 & +0.114 & -0.229 & -0.010 & -3.50 &  +4.60 & +6.42 & -0.04 \\ [4pt] \arrayrulecolor{gray}\hline
    {\it ~~Case I structure: $a$=$0.981a_0$, $b$=$0.986b_0$}&&&&&&&&& \\[-3pt]
	\multicolumn{11}{l}{\footnotesize\it (structure optimized with SOC)} \\[3pt]
    NN-Z          & \multirow{2}{*}{2.34} & 3.40 & +0.058 & +0.177 & -0.154 & -0.022 & -2.67 & -4.52 & +7.27 & -0.67  \\
    NN-X/Y      &                                        & 3.38 & +0.060 & +0.165 & -0.160 & -0.018 & -2.81 & -3.07 & +6.99 & -0.47 \\[4pt] \arrayrulecolor{gray}\hline
    {\it ~~Case II structure: $a$=$a_0$, $b$=$b_0$}&&&&&&&&& \\[-3pt]
	\multicolumn{11}{l}{\footnotesize\it (structure optimized with SOC and lattice constants fixed)} \\[3pt]
    NN-Z          &  \multirow{2}{*}{2.36} & 3.44 & +0.044 & +0.178 & -0.109 & -0.019 & -1.49 & -6.71 & +5.28 & -0.69  \\
    NN-X/Y      &                                         & 3.44 & +0.042 & +0.176 & -0.107 & -0.030 & -1.55 & -6.47 & +5.24 & -1.08  \\[4pt] \arrayrulecolor{gray}\hline
    {\it ~~Case III structure: $a$=$1.011a_0$, $b$=$1.006b_0$}&&&&&&&&& \\[-3pt]
   	\multicolumn{11}{l}{\footnotesize\it (structure optimized with SOC, $U_{\rm eff}$, and zigzag order)} \\[3pt]
    NN-Z          & \multirow{2}{*}{2.36} & 3.47 & +0.036 & +0.191 & -0.062 & -0.024 & -0.74 & -9.34 & +3.71 & -1.04  \\
    NN-X/Y      &                                        & 3.47 & +0.037 & +0.182 & -0.075 & -0.026 & -1.09 & -7.64 & +4.38 & -0.87  \\[4pt]
    \arrayrulecolor{black}\hline\hline 
  \end{tabular}
  \caption{
  Values of the averaged Ru-Cl distances, NN Ru-Ru distances, hopping integrals, 
  and examples of exchange interactions for $U$ = 3 eV and $J_{\rm H}/U$ = 0.15\cite{rau2014generic}.
  Case I to III structures were optimized with different conditions stated inside the table, while 
  Case 0 structure is from Ref. \onlinecite{RuCl3_P3112}. 
  In Case II, lattice constants are fixed to be $a_0$ and $b_0$, while in Case 
  I and III they are allowed to relax. Hopping integrals and exchange interacitons
  are shown in eV and meV units, respectively. 
  For comparison, the values of hopping integrals and exchange interactions from
  the old structure (Case 0) in Ref. \onlinecite{hskim_RuCl3} are listed. 
  }
  \label{tab:hops}
\end{table*}

Table \ref{tab:hops} shows the hopping terms from the Wannier orbitals for four crystal structures
optimized with different conditions. There are the old $P3_112$ structure\onlinecite{RuCl3_P3112} 
used in previous work, structure with internal coordinates and lattice constants optimized with SOC, 
structure with only internal coordinated optimized (fixed $a$=$a_0$ and $b$=$b_0$), and the one
optimized with SOC, $U_{\rm eff}$ and the zigzag order. Hereafter we denote the structures as 
Case 0 to III, respectively, as stated in Table \ref{tab:hops}. With those optimized structures, calculations 
of the Wannier orbitals were performed without the inclusion of SOC, $U_{\rm eff}$, and magnetism. 
Surprisingly, the hopping integrals are showing huge dependence to the structural change. 
Especially, the $t_3$ term varies from -0.229 to -0.062 eV depending on the structures, 
and $t_2$ also varies from 0.114 to 0.191 eV. 
Comparing the Case 0 and II results, the effect of Cl relaxation is to enhance $t_2$ and suppress
$t_3$. The effect of increasing Ru-Ru distance, which can be seen by comparing Case I to III, 
is also similar to the role of Cl relaxation with less dramatic but still substantial trend. 
Such tendency can be understood from the character of participating Wannier orbitals shown in Fig. \ref{fig:wan}. 
The $t_3$ term, the most sensitive to the structural change, originates from the two distinct
channels; one from the $\sigma$-like direct $d$-$d$ overlap and another from $d$-$p$-$d$ 
indirect channel. The two channels has opposite sign to each other, with minus sign for the $d$-$d$
channel and plus sign for the $d$-$p$-$d$ channel. As a result, enhancing $d$-$p$-$d$ channel 
by reducing the Ru-Cl distance or increasing the Ru-Cl-Ru angle will lead to better cancellation 
of the dominant $d$-$d$ channel and reduction of the overall $t_3$ term as shown in Table \ref{tab:hops}. 
Enhancement of $t_2$ after Cl relaxation is also easy to understand since it mostly comes from the 
$\pi$-like $d$-$p$-$d$ channel, while the $t_3$ dominated by the $\delta$-like $d$-$d$ channel
is reduced as the Ru-Ru distance is increased. The trend for the small $t_4$ term is less clear, but 
it tend to enhance when there are more trigonal and monoclinic distortion.

From the NN $t_{\rm 2g}$ hopping terms, one can estimate the values of exchange 
interaction terms in the $j_{\rm eff}=1/2$ spin Hamiltonian
\begin{equation}
\mathcal{H} = \sum_{\langle ij \rangle} {\bf S}_i \cdot {\bf M}_{ij} \cdot {\bf S}_j,
\nonumber
\end{equation}
where the bond-dependent 3$\times$3 matrix ${\bf M}_{ij}$ has the form of
\begin{equation}
{\bf M} = \left(
\begin{array}{ccc}
J & \Gamma & \Gamma' \\
\Gamma & J & \Gamma' \\
\Gamma' & \Gamma' & J+K
\end{array}
\right).
\nonumber
\end{equation}
Note that, ${\bf M}_{ij}$ undergoes simultaneous cyclic permutations of rows and 
columns depending on NN bond directions. Explicit expressions for the Heisenberg
$J$, the Kitaev $K$, and the symmetric anisotropy terms $\Gamma$ and $\Gamma'$ 
in terms of the hopping integrals, $U$, and the Hund's coupling $J_H$
are reported in Ref. \onlinecite{rau2014generic,rau_trigonal}. 
Using the values of $t_i$ listed in Table \ref{tab:hops} and setting $U$ = 3eV and $J_H/U =0.2$,
we can calculate the values of exchange interactions which are listed in Table \ref{tab:hops}. 
Note that, changing the values of $U$ and $J_H/U$ changes does not change the ratio between the
exchange interactions when $J_H/U > 0.05$. As shown in the table, among the exchange interactions, 
the Kitaev term shows dramatic change of changing sign after the Cl relaxation. This is due to
the enhancement and suppression of $t_2$ and $t_3$ terms. Increasing Ru-Ru distance gradually
enhances $K$ and reduces $J$ and $\Gamma$, so driving the system closer to the Kitaev spin liquid limit
with ferromagnetic $K$. Note that, comparing Case II and III, increasing the lattice constant by 1\% enhances
the $K$ term significantly. This implies the possibility of controlling the magnetism and realizing the Kitaev 
spin liquid phase with rather small amount of structural change such as epitaxial strain or uniaxial pressure. 
Another noticeable feature is the small but non-negligible $\Gamma'$ term
from the trigonal distortion, which can stabilize the experimentally observed zigzag order near the 
Kitaev spin liquid phase with $K<0$\cite{rau_trigonal}. 

\begin{figure}
  \centering
  \includegraphics[width=0.42 \textwidth]{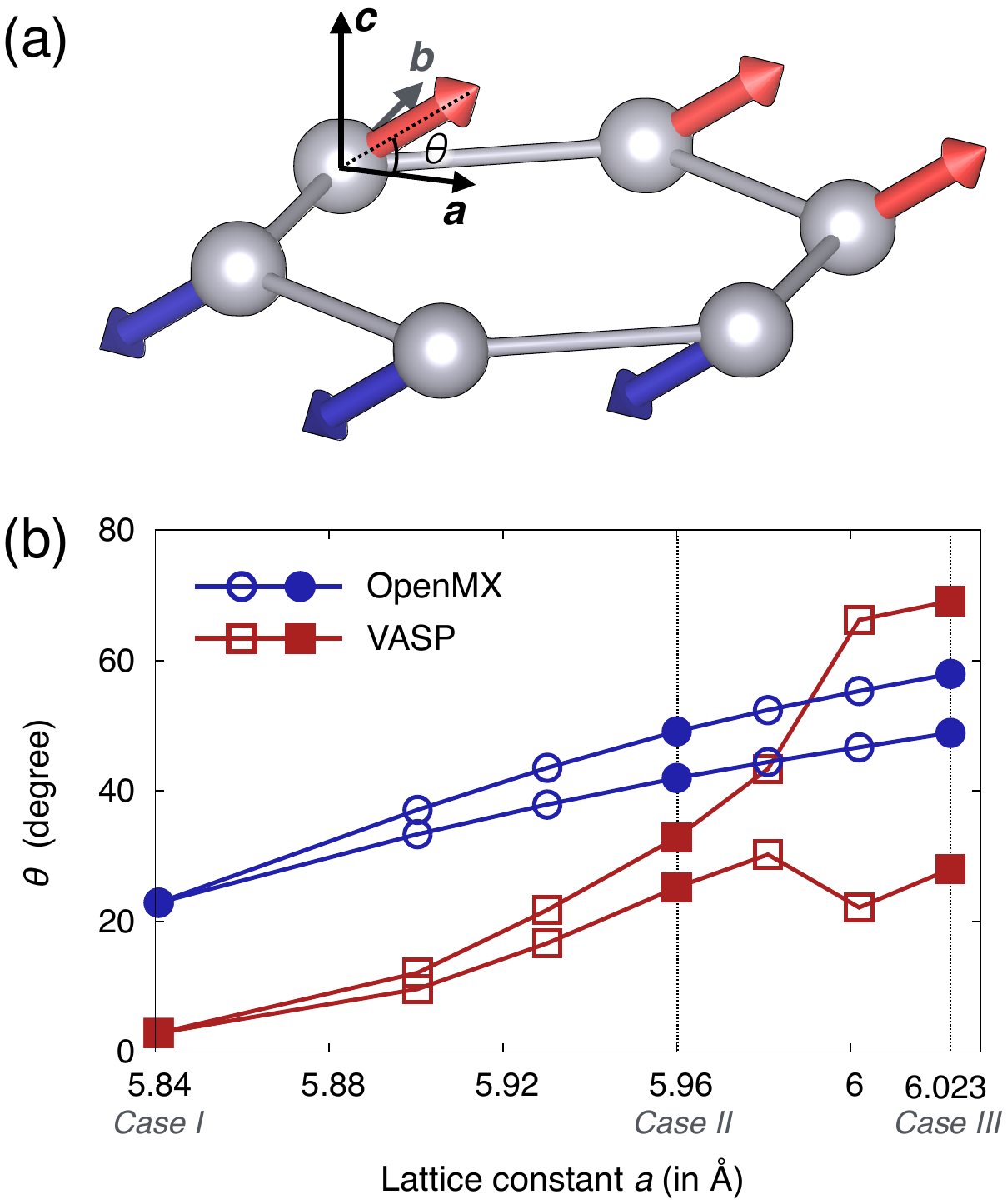}
  \caption{(Color online)
   (a) Schematic figure representing the zigzag magnetic order in the Ru honeycomb plane. 
   Note that, the moments are confined on the $ac$ plane, where the angle of the moments
   with respect to the $a$-axis is denoted as $\theta$. 
   (b) Evolution of $\theta$ for the two zigzag-ordered chains in the monoclinic unit cell
   as a function of $a$, obtained from OpenMX (blue) and VASP (red) codes. 
   Structures with $a$ = 5.84, 5.96, and 6.023\AA~ are Case I, II, and III,
   respectively, and angles from the structures are marked as filled symbols. Rest of the
   results are obtained from interpolation between the three structures, marked as empty 
   symbols. 
  }
  \label{fig:mang}
\end{figure}

Finally, we discuss the evolution of the magnetic moments direction in the zigzag order
with respect to structural changes. Fig. \ref{fig:mang}(a) shows the schematic figure of 
the zigzag order with an angle of the moments $\theta$ with respect to the $a$-axis. 
Note that, in all of our calculations the moments were residing on the $ac$-plane. 
In the Case 0 structure, both in the OpenMX and VASP results, the moments were 
parallel/antiparallel to the $a$-axis ({\it i.e.} $\theta=0$), consistent to our 
previous result\cite{hskim_RuCl3}. After the structural optimization the moments 
gain nonzero $\theta$, which tends to increase when the lattice constant increases
as shown in Fig. \ref{fig:mang}(b). There is difference in $\theta$ between the results from the
two different codes, but the tendency of increasing angle remains the same. 
We speculate that such behavior may originate from the Cl relaxation and the resulting
change of exchange interactions, especially the change of the ratio between $K$ and 
$\Gamma$ terms. Also, as the lattice constant is enlarged, the two zigzag chains with antiparallel 
moments in the unit cell begin to develop the difference in $\theta$, so resulting net 
ferromagnetic component in the $ac$-plane. The origin of such behavior is unclear at this point.

\section{Stacking without $U_{\rm eff}$ and magnetism}
\label{sec:23LwoSOC}

\begin{figure*}[htb!]
  \centering
  \includegraphics[width=0.90 \textwidth]{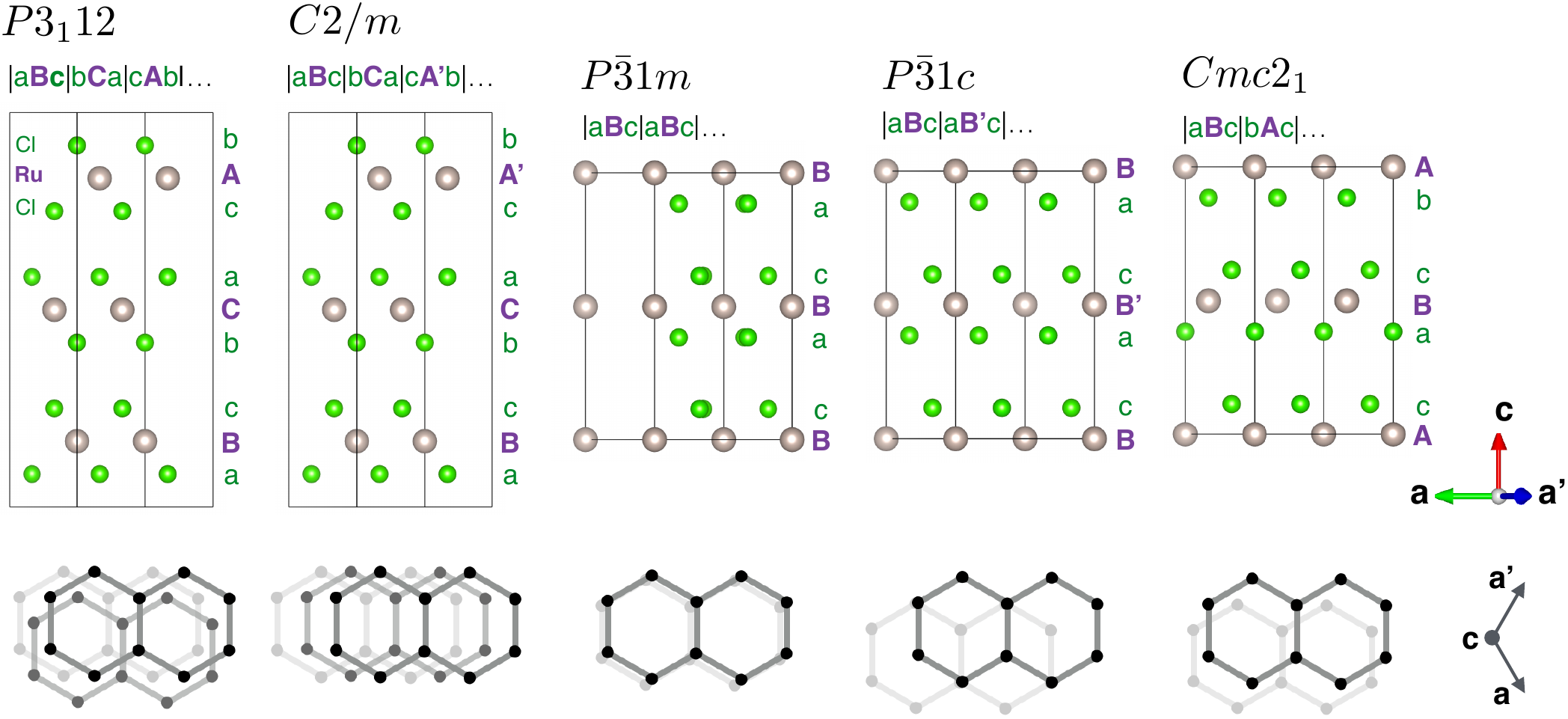}
  \caption{(Color online) 
  Five different unit cells with two- and three-layer
  periodicity along $c$-direction. Upper and lower panels
  show the side view of the unit cell and schematic top 
  view of Ru honeycomb stacking, respectively.
}
    \label{fig:stacking}
\end{figure*}

Next let us study the stacking order of RuCl$_3$. First we discuss
their relative total energies without including $U_{\rm eff}$ and magnetism. 
As mentioned in Sec. \ref{sec:detail}, here we show the results with using 
PBEsol functional, and their comparison with vdW functional
calculations are shown in the Appendix. Note that, PBEsol results give 
the same lowest energy configurations with other vdW results, and the closest
$c$-axis constant to the experimentally observed one as well\cite{Stroganov1957}. 

Fig. \ref{fig:stacking} shows five unit cells we considered in this work,
where the upper and lower panels show the side view of unit cells and top view of
Ru honeycomb layers respectively. Once we consider Ru honeycomb as a triangular layer 
by ignoring Ru hollow sites, \rucl~crystal structure can be understood as
a stacking of Ru and Cl triangular layers with three triangular sublattices 
(a/A, b/B, and c/C, where capital and lowercase letters denote Ru and Cl layers
respectively) shown in Fig. \ref{fig:struct}(a) as a degree of freedom. In 
Fig. \ref{fig:stacking}, each different structure can be understood as a 
sequence of sublattice indices. Note that, within a RuCl$_3$ layer, any two
Ru or Cl layers cannot be in a same sublattice. As we take into account Ru hollow sites,
additional degree of freedom is introduced to each Ru layer, and we denote this
with primes in the triangular sublattice index (for example, A, A', and A'' as shown
in the figure). 

For structures with three-layer $c$-axis periodicity, we choose unit cells with 
$P3_112$ and $C2/m$ space groups. Note that, the $C2/m$ structure was reported also 
as the space group of this compound\cite{RuCl3_C2m,Kubota_C2m_XY}, and is similar 
to the $P3_112$ structure. The major difference in two structures is the $c$-axis 
ordering of the Ru honeycomb layers, where in the $C2/m$ unit cell three Ru layers 
are related by translation by $({\bf a} + {\bf a}^{'} + {\bf c})/3$ while in the $P3_112$ 
cell they are related by threefold screw axis. Besides,
since the neutron diffraction result identified a magnetic peak with two-layer
$c$-axis periodicity at $T_{N1}=14$ K in a polycrystalline sample\cite{Banerjee}, we 
consider two-layered unit cells as well. Avoiding two Cl$^-$ triangular layers 
belonging to adjacent RuCl$_3$ layers to locate on top of each other ({\it i.e.} 
sitting on the same triangular sublattice), we have only three unit cells with space 
group $P\bar{3}1m$, $P\bar{3}1c$, and $Cmc2_1$ as shown in Fig. \ref{fig:stacking}. 
Note that, the $P\bar{3}1m$ cell is just a doubling of single-layer unit cell,
and the $P\bar{3}1c$ structure differ from the $P\bar{3}1m$ structure by the position 
of Ru hollow sites,
so that half of Ru sites avoid sitting on top of Ru sites in the neighboring layer
as shown in bottom panels of Fig. \ref{fig:stacking}. Finally, the $Cmc2_1$ structure 
differs from other unit cells by anti-cyclic stacking of every other RuCl$_3$ layer
as shown in the stacking sequence in the figure,
which can be obtained by applying mirror operation to every other RuCl$_3$ layers.

\begin{table}
  \centering
  \setlength\extrarowheight{2pt}
  \begin{tabular}{@{\extracolsep{\fill}}lrrrrr} 
   \hline\hline   & $P3_112$ & $C2/m$ & $P\bar{3}1m$ & $P\bar{3}1c$ & $Cmc2_1$  \\ \hline
    Lattce constants &&&&& \\
    $a/a_0$  & 0.984    & 0.981  & 0.986        & 0.985        & 0.984 \\
    $b/a_0$  & 0.984    & 0.986  & 0.986        & 0.985        & 0.983 \\
    $c/c_0$  & 1.014    & 1.013  & 1.005        & 1.007        & 1.014 \\ [5pt]
    $\Delta E$ / f.u. &&&&& \\
    (in meV) & 1.4      & 1.4    & 0.0          & 2.8          & 2.5 \\ [5pt]
    \multicolumn{2}{l}{DOS at $E_f$} &&&& \\
    (in states / eV / f.u.) & 9.2      & 7.9    & 6.0          & 10.8         & 8.5 \\    
    \hline\hline 
  \end{tabular}
  \caption{
  Optimized lattice constants, relative total energies ($\Delta$E) per formula unit (f.u.),
  and densities of states (DOS) at the Fermi level for five stacking unit cells. 
  Values are obtained using PBEsol functional and including SOC, but without electron 
  interactions.}
  \label{tab:optcoord}
\end{table}

\begin{figure}
  \centering
  \includegraphics[width=0.30 \textwidth]{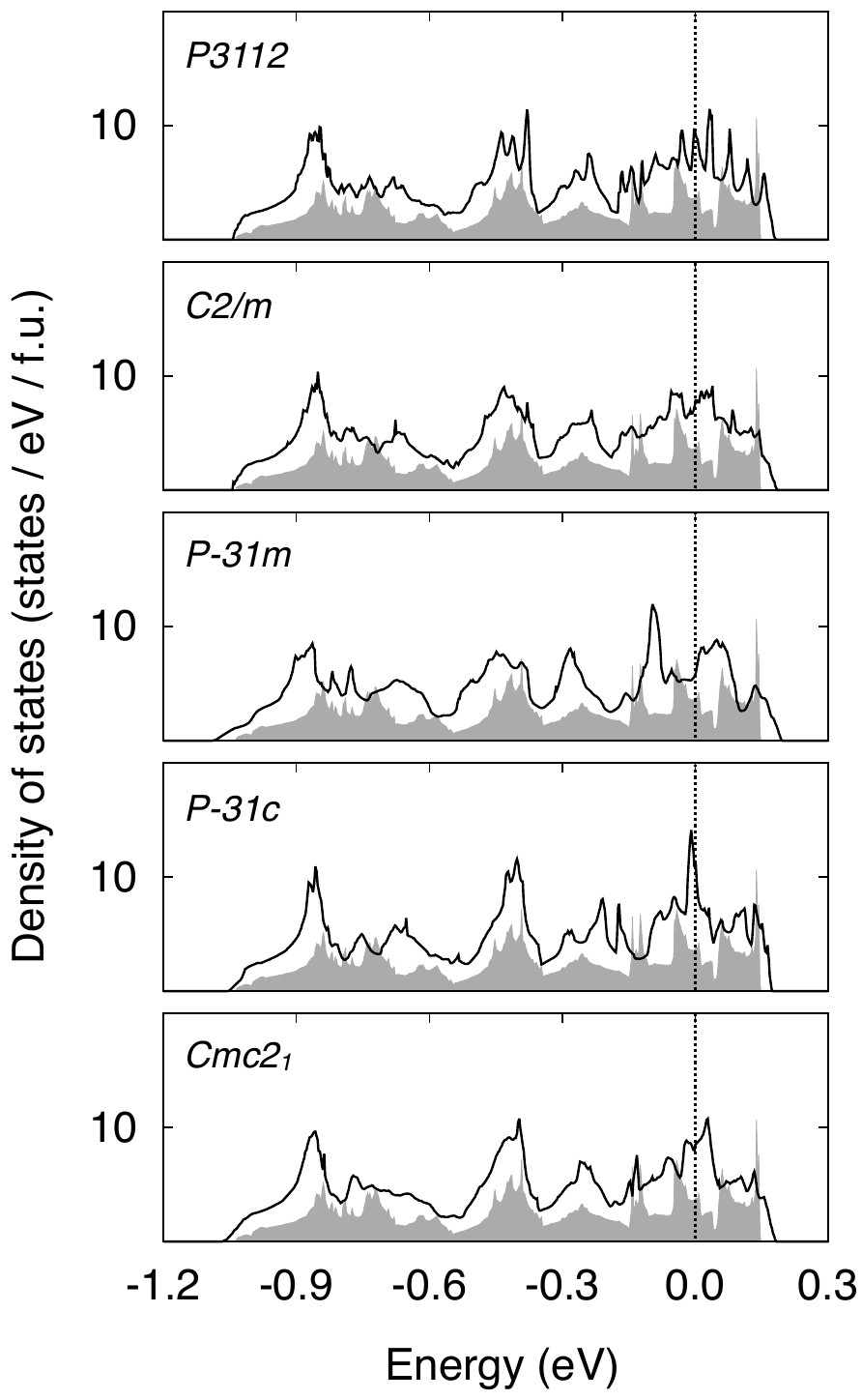}
  \caption{
  Densities of states (DOS) for different \rucl~ structures including SOC 
  in the absence of $U_{\rm eff}$ and magnetism. Grey shade shows DOS of 
  single-layer RuCl$_3$ multiplied by 0.5 as a reference. 
  }
  \label{fig:DOS}
\end{figure}

Structure optimizations were performed including SOC, and
Table \ref{tab:optcoord} shows the optimized lattice constants with respect to
experimentally reported lattice constants $a_0=5.96$\AA~and $c_0=17.2$\AA~and 
their relative total energies. Note that, structures without threefold symmetry
--- monoclinic $C2/m$ and orthorhombic $Cmc2_1$ --- shows slightly different
$a/a_0$ and $b/b_0$.
Among the five different structures, the $P\bar{3}1m$ structure yields the lowest energy.
The $P3_112$ and $C2/m$ structures are closer in energy by 1.4 meV / f.u., 
and for the other phases energy differences are less than 3 meV / f.u. compared to 
the the $P\bar{3}1m$ structure.
The lowest energy of the $P\bar{3}1m$ structure can be
attributed to the lager kinetic energy gain originating from the larger
band dispersion along the $c$-direction compared to other structures.
This is reflected in the lower DOS of the  $P\bar{3}1m$ cell at the Fermi level 
compared to other structures, as shown in Table \ref{tab:optcoord} and Fig. \ref{fig:DOS}. 
Fig. \ref{fig:DOS} presents total DOS for the five structures in the presence of
SOC. Compared to the single-layer result depicted as grey shade in the figure, 
layer stacking yields pronounced peaks near the Fermi level except the $P\bar{3}1m$ structure
in the results without SOC (not shown) due to the presence of flat bands along the
$c$-direction at the Fermi level. Inclusion of SOC smoothes the peaks,
but the gross feature remains the same as shown in Fig. \ref{fig:DOS}, so resulting 
in higher DOS at the Fermi level except the $P\bar{3}1m$ structure as shown in
Table \ref{tab:optcoord}. Note that, Stoner-type ferromagnetic (FM) instability 
is also observed, but in this study we concentrate on
the experimentally observed zigzag magnetic order as discussed in the next section.

\section{Stacking with zigzag magnetic order}
\label{sec:23LwSOCU}

\begin{figure}
  \centering
  \includegraphics[width=0.48 \textwidth]{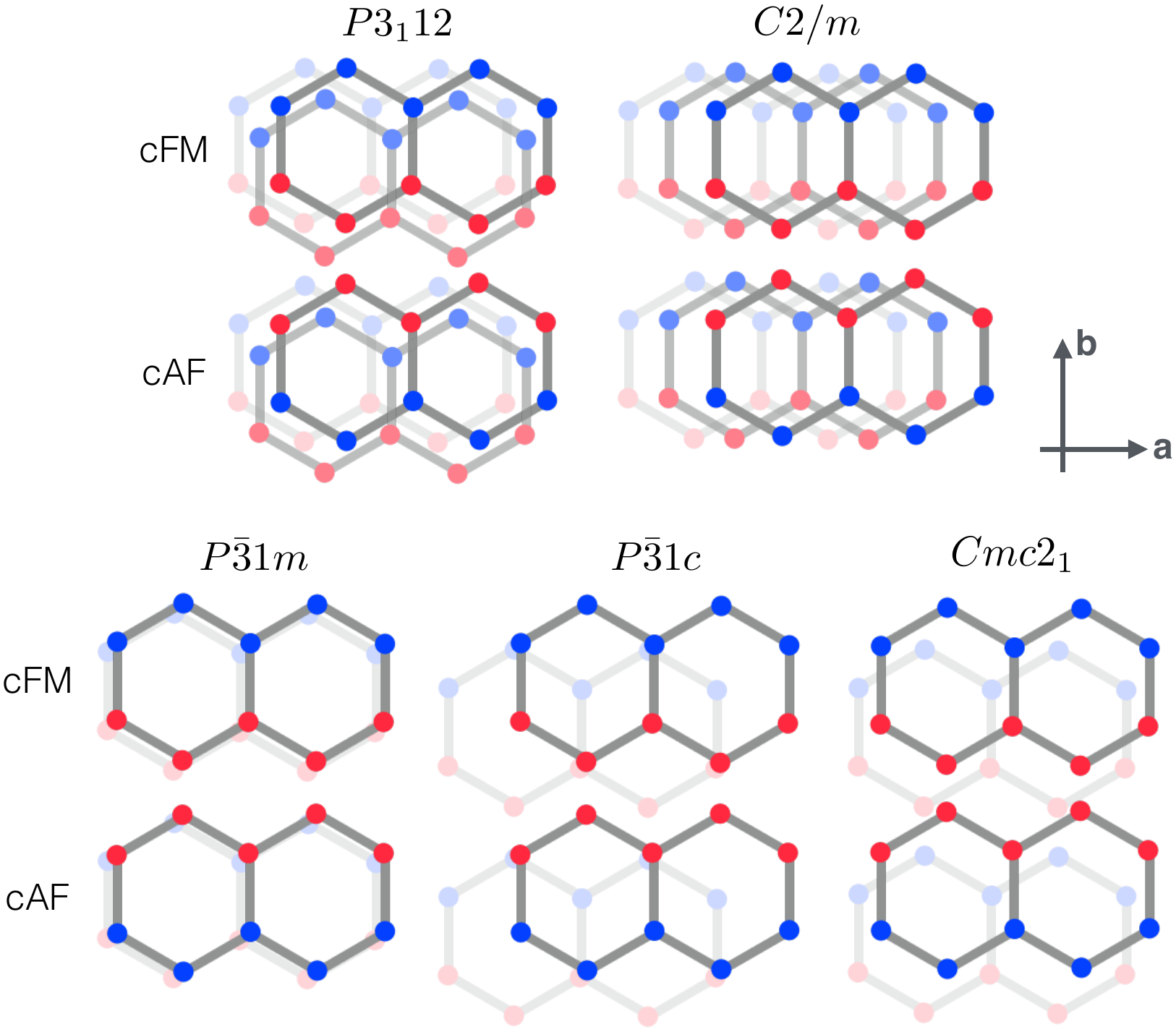}
  \caption{(Color online) 
  10 trial magnetic configurations with in-plane 
  zigzag order, where red and blue symbols depicting
  Ru sites with antiparallel magnetic moments to each other.
}
    \label{fig:mag}
\end{figure}

\begin{table}
  \centering
  \setlength\extrarowheight{2pt}
  \begin{tabular}{@{\extracolsep{\fill}}lrrrrr} 
   \hline\hline   & $P3_112$     & $C2/m$ 
                  & $P\bar{3}1m$ & $P\bar{3}1c$
                  & $Cmc2_1$  \\ \hline
    Lattice constants &&&&& \\
    $a/a_0$                  & 1.011  & 1.011  & 1.010  & 1.011  & 1.010  \\
    $b/b_0$                  & 1.006  & 1.006  & 1.006  & 1.006  & 1.006  \\
    $c/c_0$                  & 1.041  & 1.043  & 1.067  & 1.039  & 1.056  \\ [5pt]
    $\Delta E$ / f.u. (meV)  &&&&& \\ 
    cFM                      & 0.4 & 0.1 & 3.7 & 0.8 & 0.0 \\
    cAF                      & 0.4 & 0.2 & 4.1 & 0.9 & 0.4 \\
    \hline\hline 
  \end{tabular}
  \caption{
  Optimized lattice constants for five stacking unit cells with using PBEsol functional 
  and including SOC, $U_{\rm eff}$ and magnetism. $a$, $b$, and $c$ are the 
  optimized monoclinic lattice constants (shown in Fig. \ref{fig:struct}) with
  $a_0$, $b_0$, and $c_0$ being their experimentally observed values, 
  respectively\cite{Stroganov1957}. 
  }
  \label{tab:optcoordZZ}
\end{table}

Now we present the stacking results that include the on-site 
Coulomb interaction and magnetism. Fig. \ref{fig:mag} shows 10 
trial structural and magnetic configurations, where the direction 
of magnetic moments in each layer is the same with the single-layer 
result in Sec. \ref{sec:1L}. Fixing the in-plane zigzag order, we 
chose two interlayer magnetic configurations that we denote as cFM 
and cAF hereafter. As shown in Fig. \ref{fig:mag}, in the cFM configuration 
the zigzag-ordered layers are stacked along the $c$-direction so that 
the FM zigzag chains in adjacent layers become closer in distance, 
while in the cAF configuration the moments on one Ru layer are flipped.
Note that, there can be additional magnetic stacking orders due to the
threefold rotational degree of freedom for each single-layer zigzag order,
--- three different direction for FM zigzag chains ---
and in this work we chose the simplest configuration commensurate to 
the monoclinic unit cell (shown in Fig. \ref{fig:1layer}(a)) for each 
structure. Structural optimizations were done first by varying $c$-axis
with fixing $a$-lattice constants determined in the single-layer calculation,
and later fully optimizing $a$ and $c$ axis constants and internal coordinates. 
Note that, symmetry constraints are lost during the full optimizations 
including the Coulomb interaction and magnetism. As a result, the optimized 
structures slightly deviate from the original space group symmetries, where
the deviation develops in Cl positions with its size about 1\% for each
internal coordinate compared to the lattice constants. Note also that, 
structural optimization for each stacking with different magnetic 
configuration (either cFM or cAF configurations in Fig. \ref{fig:mag}) yielded 
negligible differences. All of the configurations become insulator with the gap of
$\sim$ 1 eV between the lower and upper Hubbard bands at $U_{\rm eff}=2$ eV.
DOS for the resulting phases are almost identical to the one from single-layer 
calculation\cite{luke} and show no significant difference compared to 
each other, so we do not present the DOS plots here. 

Table \ref{tab:optcoordZZ} shows the optimization results. Compared to the 
results without stacking and magnetism, a few differences can be noticed;
i) Energy differences between structures are less than 1 meV per f.u. 
except the $P\bar{3}1m$ structure, which is higher in energy by $\sim$ 4.0 meV / f.u. 
compared to other structures. Note that, the $P\bar{3}1m$ structure showed the lowest energy 
in the calculation without $U_{\rm eff}$ and magnetism. With $U_{\rm eff}$ and magnetism 
introduced, gap is fully opened for all of the structures and the relative energy gain in
the $P\bar{3}1m$ structure due to the $c$-axis dispersion 
(discussed in Sec. \ref{sec:23LwoSOC}) becomes smaller.
ii) Energy differences between cFM and cAF configurations are smaller than 0.1 meV / f.u. for 
the $P3_112$, $C2/m$, and $P\bar{3}1c$ structures, and for the $P\bar{3}1c$ and $Cmc2_1$ stackings
the differences are 
about 0.4 meV / f.u.. Such small energy differences can be attributed to weak interlayer
exchange interactions, which will be discussed later in the last paragraph of this section.
iii) Lattice constants are increased by 2 to 3 \% compared to the results without
$U_{\rm eff}$. iv) Small monoclinic distortion, which manifests itself by the difference of 
$a/a_0$ and $b/b_0$ (and negative $\delta$ in Fig. \ref{fig:1layer}), happens in every 
structures in the presence of the in-plane zigzag magnetic order.
 
Except the $P\bar{3}1m$ structure which is higher in energy by $\sim$ 4 meV / f.u. compared 
to other structures, the structural energy differences are smaller than 1 meV. 
This result implies the coexistence of different structures in experimentally synthesizes samples.
Especially, it is natural that the $P3_112$ and $C2/m$ structures have similar total energies; 
their only difference is the stacking of the Ru honeycomb order, which can be switched to each 
other by the ionic hopping of Ru atoms within the RuCl$_3$ layers. Indeed, both were reported as
the crystal structure of \rucl~by different groups\cite{Stroganov1957,RuCl3_C2m,Banerjee}. 
It is also interesting that, the $Cmc2_1$ structure (with cFM order) shows the 
lowest energy, which can be transformed into other structures by applying mirror operations to 
every other RuCl$_3$ layers. One can speculate that the $Cmc2_1$ structure forms in high temperature 
regime and freeze below $T\sim 150$ K, where an anomalous behavior in 
magnetic susceptibility observed\cite{Kubota_C2m_XY,jasears}, so contributing to the 
magnetic peak with two-layer periodicity in polycrystalline samples below $T_{N1} \simeq 14$ K
\cite{jasears,Majumder,Banerjee}. 


\begin{figure}
  \centering
  \includegraphics[width=0.3 \textwidth]{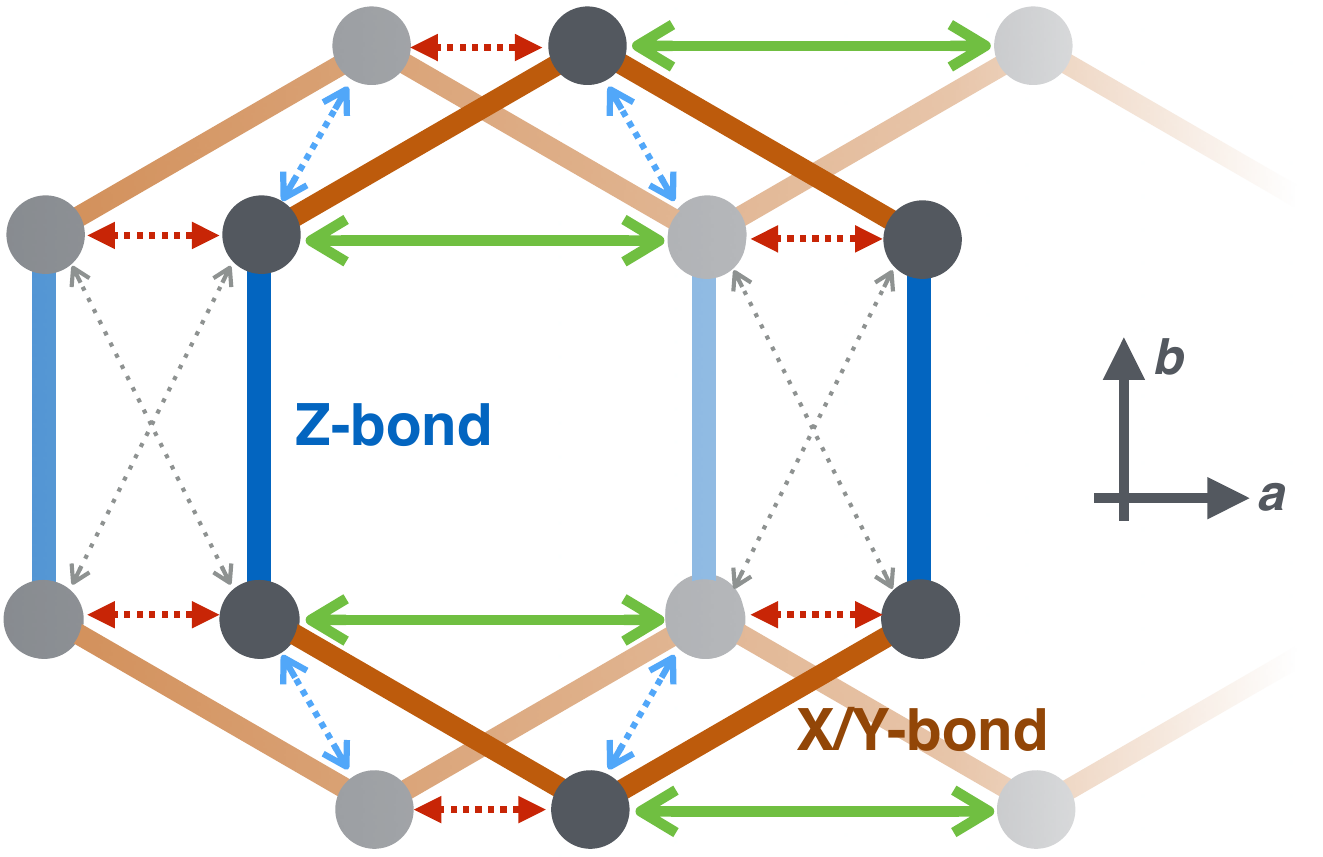}
  \caption{(Color online) 
  Two distinct NN bonds --- Z- and X/Y-bonds --- in the presence of 
  zigzag magnetic order, and the dominant interlayer hopping channels 
  in $C2/m$ structure. Note that, although the largest interlayer hopping channels 
  is depicted as green thick and solid arrows,  magnitudes of all of the hopping 
  channels in the figure are comparable to each other.
}
    \label{fig:hops}
\end{figure}

Finally, we comment on the interlayer exchange interactions.
Major interlayer hopping channels are shown in Fig. \ref{fig:hops}, 
where the largest channel is depicted as green solid arrow while others are
represented as dashed/dotted arrows. Note that, value of the largest interlayer $t_{\rm 2g}$ 
hopping term is about 35 meV, and magnitudes of other channels depicted in 
the figure are comparable to the largest one; about 20 to 30 meV. 
The interlayer exchange Heisenberg term is roughly estimated to be $J = t'^2/9U \sim 0.05$ meV 
for the $j_{\rm eff}=1/2$ pseudospins. This value is two orders-of-magnitude smaller than the 
previously estimated in-plane exchange interactions in \rucl\cite{hskim_RuCl3,Banerjee}, 
and is also consistent with the small energy differences between the cFM and cAF phases 
discussed above.

\section{Discussion} 
\label{sec:sum}

The relative energies between different stacking order depends on the electronic 
structures of each system in our results, especially whether the system becomes 
fully insulating or not. Given that \rucl~remains insulating in the 
paramagnetic phase above $T_{N1}$ with 1 eV of optical gap\cite{plumb2014alpha,luke},
we speculate that 
the four stacking orders --- $P3_112$, $C2/m$, $P\bar{3}1c$, and $Cmc2_1$ --- are 
almost degenerate as discussed in Sec. \ref{sec:23LwSOCU}. 



The change of hopping integrals and exchange interaction terms after the
structure optimization show that the physics of \rucl~is sensitive the
NN Ru-Ru distance and Cl position. 
For example, the strength of the Kitaev and $\Gamma$ terms are significantly modified 
by the Ru-Ru and Ru-Cl distances. This implies that,
even a small amount of epitaxial tensile strain by 1\% or uniaxial pressure
perpendicular to the layer can significantly enhance the Kitaev term and push the system
closer to the Kitaev limit. On the other hand, hydrostatic pressure or compressive strain
can increase the $t_3$ term by decreasing the Ru-Ru distance. This 
reduces the FM Kitaev term and drive the effective model to the highly frustrated
$\Gamma$-dominated regime. In addition, presence of the negative $\Gamma'$ term due to the trigonal 
distortion can stablize the zigzag-ordered phase as discussed 
in previous study\cite{rau_trigonal}. Effects of the monoclnic bond 
disproportionation\cite{Kimchi} is another factor that can change the magnetism. 
In this regard, full experimental structure determination including precise atomic positions 
and stacking order would be important for future studies.



In summary, structural properties of \rucl~from {\it ab-initio} calculations are
presented in this study. SOC is found to prevent the Ru dimerization in the Ru honeycomb 
layers, and the presence of in-plane zigzag magnetic order
further gives small monoclinic distortion. The relation between the hopping integrals
and exchange interactions to the structure is also discussed. 
Total energy comparison between different 
RuCl$_3$ stacking orders yields the $Cmc2_1$ and $C2/m$ structures to be the almost 
degenerate ground state structures, and $P3_112$ structure to be comparable in energy; 
energy differences smaller than 0.4 meV per formula unit.
In-plane exchange interactions are found to be sensitive to the structural distortions, 
and the $j_{\rm eff}=1/2$ pseudospin model is dominated by the FM Kitaev terms 
in the optimized structures with the presence of $U_{\rm eff}$, similar to the two- 
and three-dimensional honeycomb iridates\cite{Yamaji,Katukuri,hskim_bLIO}.
As expected, interlayer exchange interactions are estimated to be weak compared to the 
in-plane exchange interactions, so this system can be a good platform in studying 
frustrated two-dimensional magnetism. 


{\it Note added} --- After the completion of the manuscript, we became aware of the
experimental work by Johnson and co-workers\cite{Johnson-arXiv}, which reports
monoclinic $C2/m$ crystal structure and the in-plane zigzag magnetic configuration with 
antiferromagnetic interplanar order below $T_N \sim 13$ K.


\begin{acknowledgments}
We thank R. Coldea and S. Nagler for useful discussions. 
This work was supported by the NSERC of
Canada and the center for Quantum Materials at the University of
Toronto.  Computations were mainly performed on the GPC supercomputer
at the SciNet HPC Consortium. SciNet is funded by: the Canada
Foundation for Innovation under the auspices of Compute Canada; the
Government of Ontario; Ontario Research Fund - Research Excellence;
and the University of Toronto.  
\end{acknowledgments}


\appendix

\setcounter{figure}{0}
\makeatletter 
\renewcommand{\thefigure}{A\@arabic\c@figure}
\makeatother

\setcounter{table}{0}
\makeatletter 
\renewcommand{\thetable}{A\@arabic\c@figure}
\makeatother

\section{van der Waals calculation}

\begin{figure*}
  \centering
  \includegraphics[width=0.9 \textwidth]{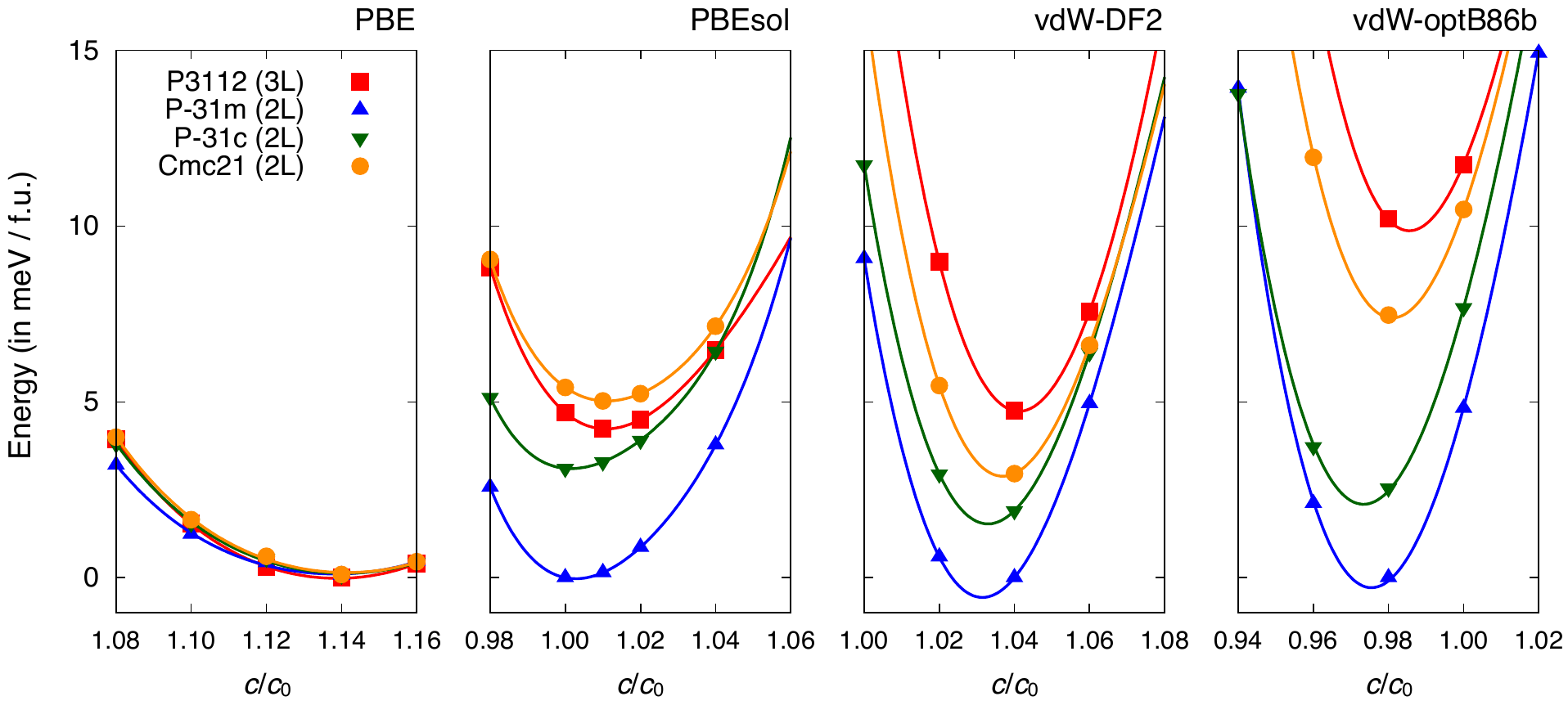}
  \caption{(Color online) 
  Total energy vs. $c$-lattice constant plots with fixed $a=a_0=5.96$\AA~for 
  different crystal structure and exchange-correlation functionals. 
  From left to right, results with PBE, PBEsol, vdW-DF2 and vdW-optB86b
  functionals are shown. 
  }
  \label{figA:stackinge}
\end{figure*}

In this Appendix we compare the results with using different exchange-correlation 
functionals including vdW interactions. Four functionals are considered; PBE, PBEsol, 
vdW-DF2\cite{vdW-DF2} and vdW-optB86b\cite{vdW-optB86b}, where vdW-DF2 and vdW-optB86b 
functionals showed accuracies comparable to RPA calculations in layered and bulk systems, 
respectively. Here SOC, $U_{\rm eff}$, and magnetism are not included.  

Fig. \ref{figA:stackinge} shows relative energies versus $c$-lattice constant 
with fixed $a=a_0$ for the results with four functionals, where $C2/m$ stacking
order is not considered. Except PBE, which yields unreasonably large value of
$c$, other three functionals yields $P\bar{3}1m$ and $P\bar{3}1c$ as configurations
with the lowest and second lowest energy. Compared to PBEsol, vdW functionals tend
to yield steeper energy curve away from the optimum $c$ valuex and higher energy for $P3_112$ phase. 

\begin{table}
  \centering
  \setlength\extrarowheight{2pt}
  \begin{tabular}{@{\extracolsep{\fill}}lrrrr} 
   \hline\hline   & $P3_112$     & $P\bar{3}1m$ 
                  & $P\bar{3}1c$ & $Cmc2_1$  \\ \hline
    $a/a_0$ &&&& \\
    PBE              & 1.011  & 1.012  & 1.006  & 1.000  \\
    PBEsol           & 0.984  & 0.986  & 0.985  & 0.984  \\
    vdW-DF2          & 1.027  & 1.029  & 1.030  & 1.027  \\ 
    vdW-optB86b      & 1.000  & 1.000  & 1.000  & 1.000  \\ [5pt]
    $c/c_0$ &&&& \\
    PBE              & 1.098  & 1.091  & 1.132  & 1.140  \\
    PBEsol           & 1.014  & 1.005  & 1.007  & 1.014  \\
    vdW-DF2          & 1.015  & 1.004  & 1.001  & 1.010  \\ 
    vdW-optB86b      & 0.980  & 0.980  & 0.980  & 0.980  \\ [5pt]
    $\Delta E$ / f.u. (meV)  &&&& \\ 
    PBE              & 1.4  & 0.0  & 2.6  & 4.5  \\
    PBEsol           & 2.1  & 0.0  & 3.0  & 2.3  \\
    vdW-DF2          &10.5  & 0.0  & 1.0  & 7.0  \\ 
    vdW-optB86b      & 9.9  & 0.0  & 2.3  & 7.1  \\ 
    \hline\hline 
  \end{tabular}
  \caption{
  Optimized lattice constants and total energy differences for four stacking 
  orders with using PBE, PBEsol, vdW-DF2 and vdW-optB86b functionals. SOC is 
  not included in these calculations. 
  }
  \label{tabA:vdW}
\end{table}

Table \ref{tabA:vdW} shows the results from full lattice optimizations.
Except the change of $a$-lattice constants, where vdW-DF2 results yields 
3\% enhancement of $a$ value, the features are qualitatively 
similar to the results in Fig. \ref{figA:stackinge}. $P\bar{3}1m$ 
is still the most favored configuration, and optimized $c$-lattice 
constants do not change significantly from the values in 
Fig. \ref{figA:stackinge}. It is notable that the vdW results give high energies
for $P3_112$ and $Cmc2_1$ phases, which were the favored phases in 
PBEsol+SOC+$U_{\rm eff}$ calculations. 

Compared to the vdW functionals, PBEsol yields reasonable estimates of total energy and 
lattice constants, although quantitative differences can be noticed. Since
test calculations on combining vdW functionals and DFT+SOC+$U$, which is crucial 
in understanding physics of RuCl$_3$, have not been done yet, in this study 
PBEsol functional is employed for the rest of the calculations. 

\bibliography{rucl3}

\end{document}